\documentclass[%
 aip,
 amsmath,amssymb,
 reprint,%
]{revtex4-1}

\usepackage{graphicx}
\usepackage{dcolumn}
\usepackage{bm}

\usepackage[utf8]{inputenc}
\usepackage[T1]{fontenc}
\usepackage{mathptmx}

\begin{document}

\preprint{AIP/123-QED}

\title{High Impedence Titanium Nitride Thin Films Grown by Reactive E-beam Evaporation}
\title{Superconducting TiN films grown by directional reactive evaporation}

\author{Raymond Mencia}
 \affiliation{Physics Department, Joint Quantum Institute, and Quantum Materials Center, University of Maryland, College Park, MD 20742}
 
 \author{Yen-Hsiang Lin}
 \affiliation{Physics Department, Joint Quantum Institute, and Quantum Materials Center, University of Maryland, College Park, MD 20742}
 
 \author{Vladimir Manucharyan}
\affiliation{Physics Department, Joint Quantum Institute, and Quantum Materials Center, University of Maryland, College Park, MD 20742}

\date{\today}

\begin{abstract}
We report a novel method of growing strongly-disordered superconducting titanium nitride (TiN) thin films by reactive electron-beam deposition. The normal state sheet resistance and superconducting critical temperature (Tc)
can be tuned by controlling the deposition pressure 
in the range of $1.1 \times 10^{-6}$ to $3.1 \times 10^{-5}~\textrm{mbar}$.
For $10~\textrm{nm}$ thick films, the sheet resistance  ($\mathrm{R_\square}$) reaches $1361 \Omega/\square$ and $\mathrm{T_c} = 0.77~ \textrm{K}$, which translates into an estimate for the sheet inductance as large as $L_{\square} = 2.4 ~\textrm{nH/}\square$. Benefiting from the directionality of reactive evaporation, we fabricated RF test devices with micron-sized dimensions using a resist mask and a lift-off process, which would be impossible with sputtering or atomic layer deposition methods. The spectroscopic measurements result in consistent sheet inductance values in two different device geometries and the quality factors ranged from Q = 300-2200. The loss is likely due to the presence of titanium oxynitride(\textrm{TiN}$_x$\textrm{O}$_y$) in the morphological composition of our films. The flexibility of the lift-off process suggest applications of reactively-evaporated TiN for making supporting structures around quantum circuits, such as readout resonators or compact on-chip filters.




\end{abstract}

\maketitle

\section{\label{sec:level1}Introduction}
High kinetic inductance of disordered superconducting films is a useful asset for device applications, including detectors, amplifiers, resonators, and qubits. \cite{Eom2012,Day2003,Peltonen2013,Leduc2010,Samkharadze2016,Chang2013,Hazard2019} These highly disordered superconducting films usually are compound materials such as titanium nitride (\textrm{TiN}), niobium nitride (\textrm{NbN}), or niobium titanium nitride (\textrm{NbTiN}). 
The two standard processes for creating highly disordered superconducting thin films are sputtering~\cite{Driessen2012,Niepce2019} and atomic layer deposition (ALD)~\cite{Baturina2007,Shearrow2018}. However, these methods are generally incompatible with depositing through a resist mask, which would be useful for fabricating devices, especially in a situation where the wafer already contains structures from the previous fabrication step. Here we explore a novel approach which utilizes reactive electron beam (e-beam) evaporation to fabricate TiN thin films. Such a process generates a directional TiN flux for deposition while the substrate is maintained at room temperature. This allows for the patterning of a device with standard e-beam lithography resist masks without additional post deposition fabrication and processing. Our growth technique produces highly disordered films whose sheet inductance values are as high as $ L_{\square} = 2.4 ~\textrm{nH/}\square$, which is larger than any reported TiN films grown by other growth methods with similar film thickness. \cite{Shearrow2018,Coumou2013,Ohya2013}   From morphology and composition analysis, we find the strong disorder of  e-beam deposited TiN films may originate from mostly amorphous-phased titanium oxynitride (\textrm{TiN}$_x$\textrm{O}$_y$) with nano-crystalline TiN embedded sparsely through out. This fabrication process may provide a alternative way to produce devices with high kinetic inductance.  

\section{\label{sec:level1}Titanium Nitride Fabrication}
Our TiN thin films are deposited by reactive e-beam evaporation on silicon-(100) oriented substrates. The substrates were prepared by sonication in acetone and isopropanol, then blown dry with nitrogen. The devices fabricated for DC transport properties were patterned with a Hall bar geometry created by a physical shadow mask.  The devices fabricated for RF measurements were patterned by electron-beam lithography using a MMA-EL13 resist mask. 
\begin{figure}
   \centering
   \includegraphics[width=0.7\linewidth]{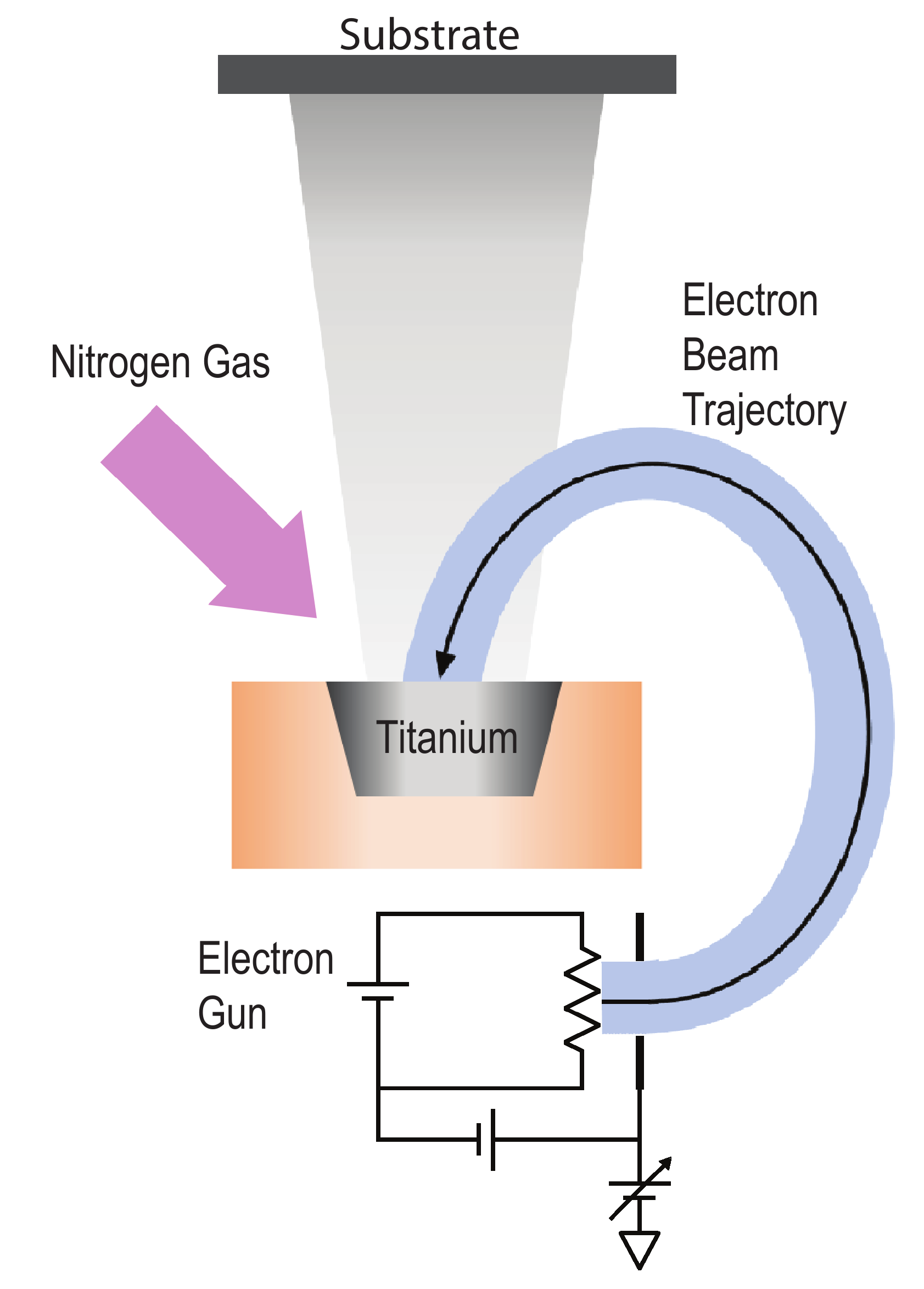}
   \caption{\label{fig:epsart} A schematic diagram of the nitrogen reactive electron beam evaporation. An ultra-high nitrogen gas flow is introduced into deposition chamber while a 10 \textrm{keV} e-beam evaporator performs titanium deposition. The precision of gas flow control is down to 0.1 cubic centimeters per minute(\textrm{sccm}).}
\end{figure}
\begin{figure*}
   \centering
   \includegraphics[width=1.0 \linewidth]{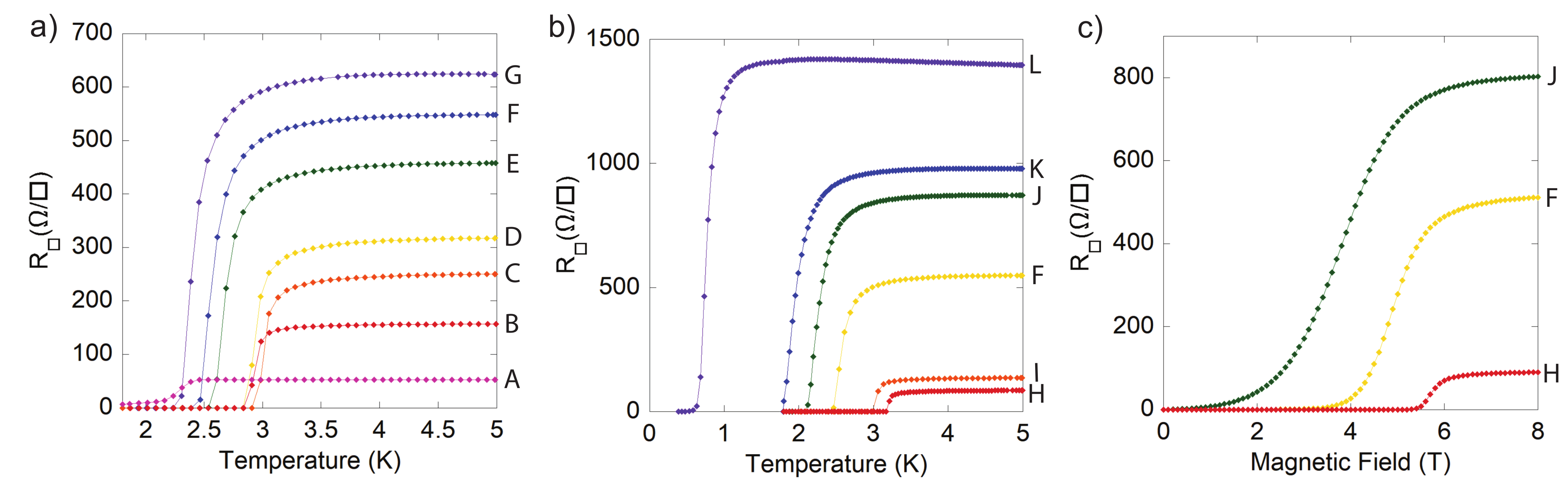}
   \caption{\label{fig:epsart} DC transport measurements of sheet resistance $R_\square $ as a function of temperature for sets of (a) samples (A to G) grown in different deposition pressures  $\mathrm{P_{dep}}$ and (b) samples (F and H to L) of different film thickness $\mathrm{d}$. And (c) $R_\square $ as function of magnetic field of sample H, F, and J. The growth condition of all samples are listed in Table 1.}
\end{figure*}
The substrates were then loaded into a Plassys MEB550S E-beam evaporation system where the main deposition scheme is shown in FIG.1. The main deposition chamber is pumped down to a pressure below $5.0 \times 10^{-7}$\textrm{mbar} before beginning the  deposition. The deposition procedure started with heating a titanium (with purity 99.995\% ) source with the e-beam. Once the titanium evaporation rate stabilized, a controlled flow of ultra-high pure nitrogen gas (with purity 99.999\%) was introduced into the deposition chamber. Due to the high energy of the e-beam and the high temperatures of the titanium source, nitrogen gas reacts with the titanium flux to form TiN. \\
During deposition, the chamber deposition pressure is determined by titanium deposition rate and nitrogen gas flow rate. Clearly, the higher the nitrogen gas flow rate, the higher the chamber pressure. However, the evaporated titanium material can both simultaneously generate titanium flux (raising the chamber pressure) and serve as a titanium sublimation pump (reducing the chamber pressure). To control the chamber pressure and reaction condition, a feedback circuit loop of the e-beam current maintains a constant deposition titanium rate at $0.15~\textrm{nm/sec}$ for all TiN films in this study. The chamber pressure is then tuned by nitrogen flow rate, which is controlled in the range of 0-20 cubic centimeters per minute (\textrm{sccm}) and has a precision of $0.1~\textrm{sccm}$.  The actual TiN deposition is only started once both deposition rate and chamber pressures have reached stable values. Unlike ALD growth, the substrate is always maintained at room temperature during the whole deposition process. Also, the evaporated material flux is directional -- same as the normal e-beam evaporation process. Thus, this film growth process is suitable for both photo- and e-beam resist mask fabrication techniques. The only required post deposition step of fabricating RF devices is the standard lift-off procedure of the e-beam mask by a heated acetone bath for approximately one hour. 
\begin{table}
\caption{\label{tab:table1}Summary of TiN film growth conditions and DC transport measurements. All films are grown with the same growth rate of $0.15~\textrm{nm/sec}$. The sheet kinetic inductance $L_\square$ is estimated by Ambegaokar-Baratoff relation and BCS theory.}
\begin{ruledtabular}
\begin{tabular}{cccccc}
Sample&$\mathrm{d}$(\textrm{nm})&$\mathrm{P_{dep}}$(\textrm{mbar})&$R_\square @10K (\Omega/\square)$&$\mathrm{T_c}(\textrm{K})$&$L_\square(\textrm{pH}/\square)$\\
\hline
A & 100 & $1.1 \times 10^{-6}$ & 52 & 2.65 & 27\\
B & 100 & $2.6 \times 10^{-6}$ & 157 & 2.95 & 73\\
C & 100 & $5.5 \times 10^{-6}$ & 250 & 3.03 & 114\\
D & 100 & $6.1 \times 10^{-6}$ & 316 & 2.95 & 148\\
E & 100 & $7.3 \times 10^{-6}$ & 451 & 2.7 & 231\\
F & 100 & $1.2 \times 10^{-5}$ & 534 & 2.58 & 286\\
G & 100 & $3.1 \times 10^{-5}$ & 600 & 2.35 & 353\\
H & 300 & $1.2 \times 10^{-5}$ & 85 & 3.17 & 37\\
I & 200 & $1.2 \times 10^{-5}$ & 135 & 3.02 & 62\\
J & 30 & $1.2 \times 10^{-5}$ & 855 & 2.4 & 492\\
K & 20 & $1.2 \times 10^{-5}$ & 961 & 1.91 & 674\\
L & 10 & $1.2 \times 10^{-5}$ & 1361 & 0.77 & 2442\\

\end{tabular}
\end{ruledtabular}
\end{table}
\section{\label{sec:level1}DC Transport Measurements}
We utilize a physical properties measurement system (PPMS) to characterize DC transport properties of TiN films. All the samples for DC transport are performed by standard four terminal measurements and summarized in Table 1. We found the deposition pressure $\mathrm{P_{dep}}$ and film thickness $\mathrm{d}$ significantly affects the properties of the grown TiN films.

\begin{figure*}
   \centering
   \includegraphics[width=0.8\linewidth]{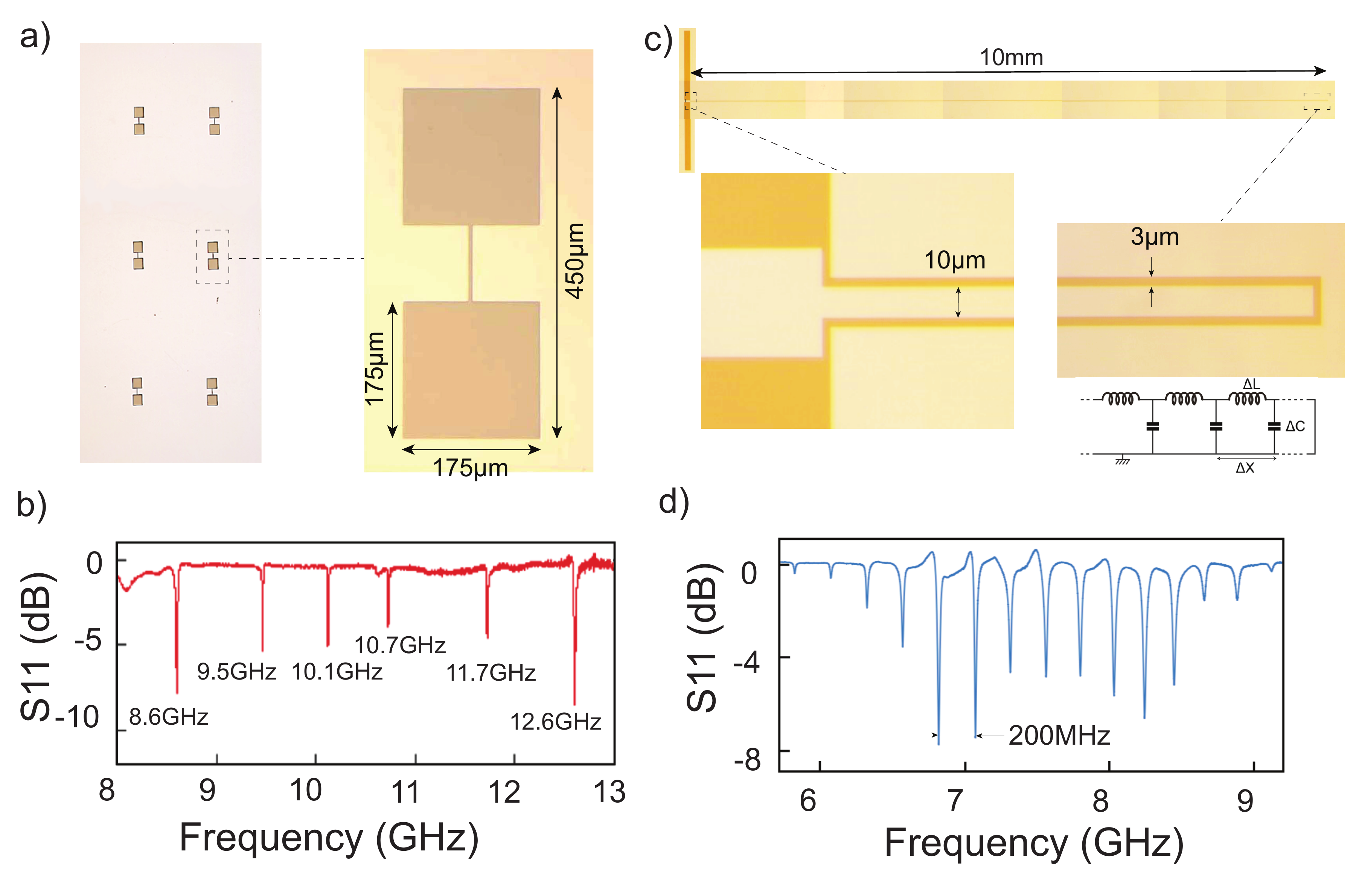}
   \caption{\label{fig:epsart} The optical microscope image of (a) six RF resonators and (c) transmission line RF device with an effective circuit model shown at the bottom. In the resonators device, the two square electrode pads provide capacitance of the resonator and couple to a 3D waveguide for measurements. The transmission line device consists of $10~\textrm{mm}$ long TiN wires. One end of the transmission line is shorted while the other end links to an antenna. The reflection response $S11$ , measured by a VNA, are shown in (b) and (d).}
\end{figure*}

In FIG.2(a), we compare sheet resistance ($\mathrm{R_\square}$) versus temperature of seven $100~\textrm{nm}$ thick films (A through G) grown at deposition pressures, $\mathrm{P_{dep}}$ , in a range from $1.1 \times 10^{-6}~\textrm{mbar}$ to  $3.1 \times 10^{-5}~\textrm{mbar}$. The normal $\mathrm{R_\square}$ at $10~\textrm{K}$ increases more than one order of magnitude from $52$ to $600 \Omega/\square$ (corresponding to resistivity from $520$ to $6000 \mu\Omega \cdot $cm, respectively). On the other hand, the $\mathrm{T_c}$ first increases but then decreases with increasing $\mathrm{P_{dep}}$. The highest $\mathrm{T_c}$ peaked at $3.03~\textrm{K}$ with the $\mathrm{P_{dep}} = 5.5 \times 10^{-6}~\textrm{mbar}$. The non-monotonic behavior of $\mathrm{T_c}$ with $\mathrm{P_{dep}}$ suggests there is a competition between  nitrogen incorporation\cite{Ohya2013,VISSERS2013} and suppression of $\mathrm{T_c}$ due to disorder.\cite{Haviland1989,Baturina2007} According to the Ambegaokar-Baratoff relation and BCS theory, the sheet kinetic inductance can be estimated as $L_\square = \hbar R_\square/1.76\pi k_B T_c $.\cite{tinkham2004} Accordingly, we obtained a wide range of  $L_\square$ for samples A to G ranging from $27~\textrm{pH/}\square$ to $353~\textrm{pH}/\square$. 

The TiN film properties are also tunable by changing the film thickness $\mathrm{d}$. FIG.2(b) shows $R_\square$ versus temperature of six films (sample F and H to L), all grown at the same $\mathrm{P_{dep}} =1.2 \times 10^{-5}$ \textrm{mbar}, with the thicknesses ranging from 10 \textrm{nm} to 300 \textrm{nm} thick. The normal $R_\square $ increases with decreasing film thickness. Interestingly, the critical temperature of TiN films decreases with decreasing film thickness. The $10~\textrm{nm}$ film shows a critical temperature of $0.77~\textrm{K}$ while it has the highest normal $R_\square = 1361 \Omega/\square$ at $10~\textrm{K}$. Such behaviors have been observed with strongly disordered superconducting films near thickness tuned superconductor-insulator (SI) transitions.\cite{Haviland1989,Baturina2007}  With such tunability, the $L_\square$ of 10\textrm{nm} TiN film reach up to 2.4\textrm{nH}/$\square$, which is nearly one order of magnitude larger than $L_\square$ of TiN films grown by sputter and ALD.\cite{Coumou2013,Ohya2013,Shearrow2018}

Additionally, the superconductivity of these TiN films can tolerate large perpendicular magnetic fields. FIG.2(c) shows the $\mathrm{R_\square}$ of sample H, F, J versus perpendicular magnetic fields at $1.8~\textrm{K}$. The critical magnetic field for sample H, F, and J are $5.3~\textrm{T}$, $2.9~\textrm{T}$, and $0.4~\textrm{T}$ respectively. The critical field is also one to two orders of magnitude larger than the typical aluminum based Josephson junction array devices. This demonstrates disordered TiN films can still serve as a high impedance device in a high magnetic field environment. 

\section{\label{sec:level1}RF Device and Measurements}
To probe the RF properties of TiN films, we patterned two different types of devices: (1) resonators (shown in FIG.3(a)) and (2) high impedance transmission lines (shown in FIG.3(c)). The measurement utilized the same setup of Kuzmin et al.\cite{Kuzmin2019}. The devices are capacitively coupled to a single-port 3D copper waveguide which is then mounted to a dilution refrigerator and the microwave reflection response is probed with a vector network analyzer (VNA).\

\begin{figure*}
   \centering
   \includegraphics[width=0.8 \linewidth]{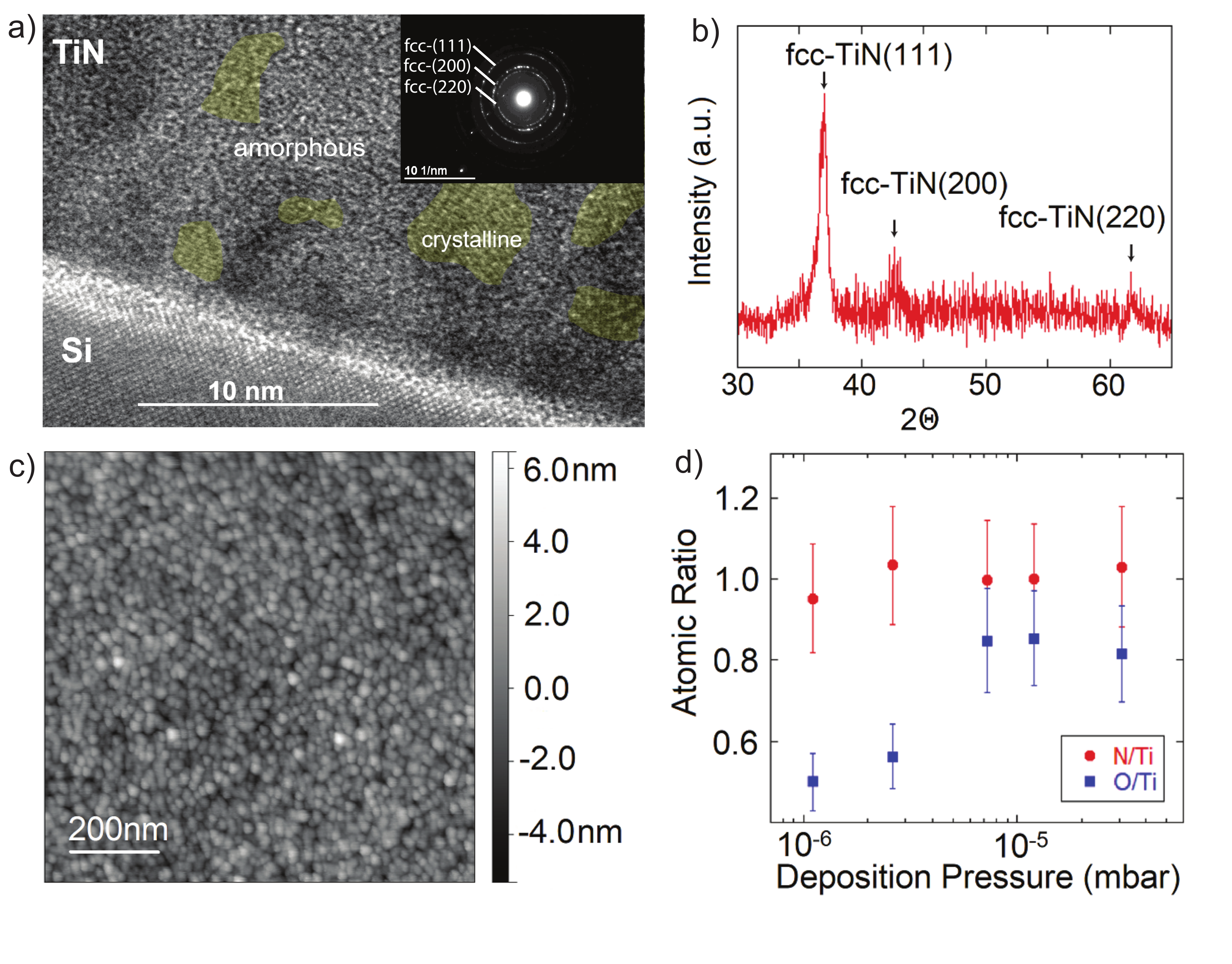}
   \caption{\label{fig:epsart} The morphology analysis of grown TiN films. (a) Transmission electron microscopy(TEM) of TiN sample grown in conditions similar to sample F. The false color area labels the crystalline features while the other area is mainly amorphous. The inset shows the selected area diffraction (SAD) pattern which reveals the crystalline orientations of TiN. (b) X-ray diffraction (XRD) of theta-2 theta analysis. (c) Atomic force microscope (AFM) scan of the surface structure. (d) Atomic ratios of nitrogen to titanium and oxygen to titanium measured by energy-dispersive X-ray spectroscopy (EDX). These results reveal the atomic composition of grown films are, in fact, titanium oxynitride \textrm{TiN}$_x$\textrm{O}$_y$.         }
\end{figure*}

The resonator device was deposited at $\mathrm{P_{dep}} = 6.3 \times 10^{-6}$ \textrm{mbar} with a 100 nm thick TiN film. There were a total  of six resonators with a separation of at least $2~\textrm{mm}$ to the nearest neighboring resonator such that the coupling between resonators is weak. The capacitance value is dominated by the two $175 \mu\/\textrm{m} \times 175\mu\/\textrm{m}$ square pads. The capacitance of the antenna was determined by HFSS simulations to be 39 \textrm{fF}. The inductance value is dominated by the kinetic inductance of the $100\mu\/m$ long thin wire connecting the two pads. The width of the wires are chosen to be $1.575 \mu\/\textrm{m}$, $2 \mu\/\textrm{m}$, $2.25 \mu\/\textrm{m}$,  $2.5 \mu\/\textrm{m}$, $3 \mu\/\textrm{m}$, and $3.55 \mu\/\textrm{m}$ for these six different resonators.

The magnitude of the reflected signal, $S11$, is shown in FIG3.(b). There are six dips at $8.6 ~\textrm{GHz}$, $9.5 ~\textrm{GHz}$, $10.1 ~\textrm{GHz}$, $10.7 ~\textrm{GHz}$, $11.7 ~\textrm{GHz}$, and $12.5 ~\textrm{GHz}$, which correspond to the resonance frequencies of the six resonators. The total inductance of each resonator can be extrapolated from the measured resonance frequencies and simulated capacitance value.  Assuming the inductance values here are all provided by the kinetic inductance of the disordered TiN film, we find the sheet inductance value of each resonator to be $139~\textrm{pH}$, $146~\textrm{pH}$, $145~\textrm{pH}$, $142~\textrm{pH}$, $145~\textrm{pH}$, and $146~\textrm{pH}$. The maximum difference between individual resonator's  sheet inductance is about 4 percent, which reveals the non-uniformity of the TiN film within a single deposition. Moreover, the thickness and deposition pressure of this device is controlled to be the same as film D in the DC measurement experiment but each sample was deposited in two different depositions. The DC measurement analysis of film D revealed a sheet inductance of $148 \textrm{pH}$, which has only a 3 percent difference to the average sheet inductance value of the six resonators in the RF measurement. The systematic difference from deposition to deposition is comparable to sputtering TiN and Josephson junction chains.\cite{Ohya2013,VISSERS2013,Masluk2012} To extrapolate the intrinsic quality factor ($Q_\textrm{int}$), we used the common expression to fit the reflection coefficient as a function of frequency:\cite{Kuzmin2019}
\begin{equation}
    S_\textrm{11}(f)=\frac{2i(f-f_0)/f_0-Q_\textrm{ext}^{-1}+Q_\textrm{int}^{-1}}{2i(f-f_0)/f_0+Q_\textrm{ext}^{-1}+Q_\textrm{int}^{-1}}
\end{equation}
We obtained $Q_\textrm{int}$ values in the range of 1500 - 2200 for the six resonance peaks.                   

The transmission line device is designed with two parallel 10 mm long, $3\mu\/m$ wide TiN wires shown in FIG.3(c). The TiN waveguide was deposited 30 nm thick with $\mathrm{P_{dep}} = 7.0 \times 10^{-6}~\textrm{mbar}$. One end of the wire is short circuited and the other end is connected to an antenna which capacitively couples to a 3D copper waveguide.\cite{Kuzmin2019}. The magnitude of single-tone reflection signal, S11, as a function of probe frequency is shown in Fig3(d). The S11 reveals resonance dips with equal frequency spacing $f_{n+1} - f_{n} = 200\textrm{MHz}$. The wave-number difference of adjacent modes is defined as $k_{n+1} - k_{n} = \pi/l$, where $l = 10~\textrm{mm}$ is the length of the line. In the measured frequency range, we observed a linear dispersion relation, which gives a slow wave velocity $v =4.0\times 10^6~\textrm{m/s}$. The value of capacitance per micrometer is $42~\textrm{aF}/\mu\textrm{m}$, which is calculated with a common formula for two coplanar strip lines on top of a silicon substrate.\cite{Fouad1980} Thus, we can obtain the sheet inductance value of TiN for this particular device as $465~\textrm{pH}/\square$. The sheet inductance value is smaller than the value of sample K found via DC measurements. Presumably this is due to a slightly lower TiN deposition pressure of the transmission line device. We also obtained the $Q_\textrm{int}$ for each mode with Eq.1. The value of $Q_\textrm{int}$ is between $300$ - $700$ with an average of $470$. 

\section{\label{sec:level1}Morphology and Atomic Composition Characterization}

To further understand the origin of the disorder and the  morphology in the TiN films, we performed various morphology analysis to the films grown with the same conditions of sample F in Table 1.

First, we used transmission electron microscopy (TEM) to analyze local crystalline morphology of the TiN films. The most striking feature is that most areas are amorphous with only sporadic poly-crystalline embedded within, shown in FIG.4(a). The false color area labeled 'crystalline area' shows the formation of nano-crystals, which typically have a size less than $5~\textrm{nm}$. The fact that the majority of the films morphology is amorphous, confirms that these TiN films are strongly disordered. The different crystalline orientations seen in TEM is determined with the selected area diffraction (SAD) pattern, shown in the the inlet of the FIG.4(a). The locations of the ring like features indicate that the different crystalline orientation are fcc-TiN (111), fcc-TiN (200), and fcc-TiN (220). \cite{JU2019} Furthermore, we performed  X-ray diffraction (XRD) theta-2 theta analysis to confirm crystalline orientations, shown in FIG.4(b).  The peaks found at $36.5^{\circ}$, $42.5^{\circ}$, and $42^{\circ}$ correspond to fcc-TiN (111), fcc-TiN (200), and fcc-TiN (220), respectively, and are consistent with SAD's results. Despite the different growth method, the same crystalline orientations have also been found in ALD and sputtered TiN thin films.\cite{Shearrow2018,VISSERS2013,Ohya2013,VAZ2002}

We then utilize atomic force microscopy (AFM) to study the surface morphology of the TiN films. FIG.4(c) shows an example of a $1\mu\textrm{m} \times 1\mu\textrm{m}$ AFM scan performed on a 100 nm thick TiN film grown at the deposition pressure of sample E. The surface of TiN consists of grains with diameter around $20~\textrm{nm}$. The root mean square surface roughness is $1.2~\textrm{nm}$ while the maximum thickness variation is less than $12~\textrm{nm}$. Therefore, films thinner than $10~\textrm{nm}$ may result in physically disconnected structures. To avoid weak links or unwanted vortex structure, the thickness of TiN films should be thicker than $30~\textrm{nm}$ while fabricating RF devices.  
 
The atomic composition of the TiN films were analyzed by energy-dispersive X-ray spectroscopy (EDX) with a $5~\textrm{KeV}$ accelerating voltage variable pressure Hitachi scanning electron microscope. Contrary to the expected composition of just titanium and nitrogen, a large amount of oxygen was present in all TiN films. FIG.4(d) shows a summary of the EDX atomic ratio of nitrogen to titanium (N/Ti) and oxygen to titanium (O/Ti) as a function of deposition pressure. The nitrogen to titanium ratio is nearly 1:1 within the measurement error, which indicates that the nitrogen atom indeed incorporates in the form of titanium nitride. Interestingly, the oxygen to titanium atomic ratio increases from $0.5$ to $0.8$ with an increase in the deposition pressure. During deposition, the partial pressure of oxygen is at least two orders of magnitude lower than nitrogen. Such large amounts of oxygen composition presumably is formed after being exposed to ambient conditions. A similar oxidation process under ambient conditions and the large amounts of oxygen in the chemical composition have also been reported in TiN thin films grown by ALD and sputtering.\cite{Shearrow2018,Ohya2013} Since SAD and XRD both confirmed that crystalline regimes are formed by TiN, we  conclude that the oxygen is diffused into the amorphous regions and form \textrm{TiN}$_x$\textrm{O}$_y$. The large portion of amorphous \textrm{TiN}$_x$\textrm{O}$_y$ reveal the origin of the strongly disordered, highly resistive properties and may also explain the low quality factors measured during our RF experiments.  

\section{\label{sec:level1}Conclusion}

In conclusion, the values of the sheet kinetic inductance of TiN films prepared by nitrogen assisted reactive e-beam deposition can be tuned by two orders of magnitude, from $27\textrm{pH/}\square$ to $2.4\textrm{nH/}\square$. The tuning knobs are deposition pressure and film thickness. The variations of kinetic inductance within the same deposition and between different depositions is within about 5\textrm{\%}. Although the quality factors of our lift-off devices is short of the values accessible with more traditional sputtering or ALD film growth methods, our process can be useful in creating compact high-impedance resonators and filters that survive in a relatively high magnetic fields. 

The method for directional reactive TiN deposition was suggested by late Patrick Smuteck to whom this article is dedicated. The authors thank Dr. Sz-Chian Liou from Advance Imaging and Microscopy Lab in University of Maryland for his assistance in TEM imaging and analysis and Dr. Joshua Higgins for assistance with XRD analysis and PPMS measurements. This work was supported by the NSF Career grant (DMR 1455261) and by ARO-LPS program “New and Emerging Qubit Science and Technology” (W911NF1810115).

\nocite{*}
\bibliography{aipsampTiN}

\end{document}